\documentclass[a4paper,aps,preprintnumbers,showpacs,twocolumn,superscriptaddress,nofootinbib,amsmath,amssymb]{revtex4-1}
\usepackage[dvips]{graphicx}

\newcommand{\rem}[1]{}
\begin{document}
\title{Analytical quasinormal modes of spherically symmetric black holes in the eikonal regime}
\author{M. S. Churilova}
\email{wwrttye@gmail.com}
\affiliation{Dnipro National University, Dnipro 49010, Ukraine}
\begin{abstract}
Quasinormal modes in the high frequency (eikonal) regime can be obtained analytically as the Mashhoon-Will-Schiutz WKB formula is exact in this case. This regime is interesting because of the correspondence between eikonal quasinormal modes and null geodesics, as well as due to  existence of potential eikonal instabilities in some theories of gravity. At the same time in a number of studies devoted to quasinormal modes of spherically symmetric black holes this opportunity was omitted. Here we find analytical quasinormal modes of black holes in various alternative and extended theories of gravity in the form of the Schwarzschld eikonal quasinormal modes and added corrections due to  deviations from Einstein theory. We also deduce a generic formula for analytical calculations of the eikonal quasinormal modes for the class of asymptotically flat metrics in terms of small deviations from the Schwarzschild geometry.
\end{abstract}

\maketitle

\section{Introduction}

There exist a number of alternative gravitational theories representing attempts to solve fundamental problems, such as the nature of the gravitational singularity, dark matter and dark energy problems, quantum gravitational theory, hierarchy problem. Recent measurements, either gravitational or electromagnetic, do not identify geometry of black holes or, possibly, other compact objects with sufficient accuracy \cite{Abbott:2016blz, TheLIGOScientific:2016src, Konoplya:2016pmh, Goddi:2017pfy, Akiyama:2019cqa, Akiyama:2019bqs, Bambi:2015kza, Cardoso:2016rao, Yunes:2016jcc, Bambi:2019tjh}.

That is why we consider different black-hole solutions of modified or extended Einstein equations. All these metrics have one thing in common: they have a small parameter or a number of parameters such that when they vanish, the metric tends to the Schwarzschild one. Therefore we can expand the metric in terms of these small parameters and do the same to find the quasinormal modes.

Quasinormal modes \cite{QNMreviews} being the source of information about the black holes have been studied in the numerous researches, but an overwhelming part of these studies is concerned with only numerical calculation of QNMs. Nevertheless in geometrical optics approximation (with high multipole number $\ell$) an elegant analytical formula was obtained  by B. Mashhoon in his seminal work \cite{MashhoonBlome}:
\begin{equation}\label{Mashhoon}
\omega=\frac{1}{3\sqrt{3}M}\left(\left(\ell+\frac{1}{2}\right)-i\left(n+\frac{1}{2}\right)\right)+ {\cal O}\left(\frac{1}{\ell}\right),
\end{equation}
where $n$ is the overtone number. A recent review on further extensions and accuracy of the WKB approach can be found in \cite{Konoplya:2019hlu}.

This formula (\ref{Mashhoon}) was extended to the Reissner-Nordstrom case in \cite{Andersson:1996xw}, to non-extremal Schwarzschild - de Sitter in \cite{Zhidenko:2003wq} and to the near extremal Schwarzschild - de Sitter case in \cite{Cardoso:2003sw}. Further extensions to the case of Kerr black hole were done in a number of papers for fields of various spin \cite{Mashhoon, Konoplya:2017tvu, Hod:2012bw}. A number of extensions of the formula (\ref{Mashhoon}) to higher dimensional spacetimes were suggested in \cite{Konoplya:2003ii, Molina:2003ff, Konoplya:2004xx, Vanzo:2004fy}. Recently eikonal regime has been studied in Einstein-dilaton-Gauss-Bonnet \cite{Konoplya:2019hml} and Einstein-Weyl \cite{Kokkotas:2017zwt} gravities as well as for the case of non-linear (derivative) coupling between scalar and electromagnetic fields \cite{Konoplya:2018qov}. The eikonal regime for black holes in non-linear electrodynamics and regular black holes have been recently studied in \cite{Toshmatov:2015wga}. However the final analytical formulae presented there are not expressed in terms of the parameters of the black holes only (such as mass, charge etc.). They also contain the maximum of the effective potential as a parameter. The summary of
 analytical results for the quasinormal modes emitted by the black hole in the eikonal regime are reviewed in Table 1.

The eikonal regime of quasinormal frequencies have been recently discussed because of the
correspondence between eikonal quasinormal modes and null geodesics \cite{Cardoso:2008bp}. The correspondence states that the real and imaginary parts of the quasinormal mode are multiples of the frequency and instability timescale of the circular null geodesics respectively. In  \cite{Konoplya:2017wot} it was shown that the correspondence is guaranteed for any stationary, spherically symmetric, asymptotically flat black hole provided that a) effective potential has the form of the potential barrier with a single extremum, implying two turning points and decaying at the event horizon and infinity and b) one is limited by perturbations of the test fields only, and not of the gravitational field itself or other fields, which are non-minimally coupled to gravity.

Another reason to study quasinormal modes in the eikonal regime is possibility of the so-called eikonal instability which may happen either for black branes \cite{Takahashi:2011du} and holes \cite{Konoplya:2017lhs} or even for wormholes \cite{Cuyubamba:2018jdl}. Eikonal instability means the breakdown of the linear approximation as such.

Here we will consider a bunch of theories in which the analytical eikonal formula has not been deduced. In particular, we shall consider two different kinds of Lorentz-violating Einstein-aether theory, Horava-Lifshitz and Horndeski theories and regular BH theories: Bardeen and Ayon-Beato-Garcia black holes in theories coupled to a non-linear electrodynamics and Hayward regular black hole solution. Finally, we develop a general formula for finding eikonal quasinormal modes of asymptotically flat black holes whose geometry slightly deviates from the Schwarzschild one.

\begin{center}
\begin{table*}
\begin{tabular}{p{10cm}p{6cm}}
\multicolumn{2}{c}{{\bf Table 1.} Analytical results for the eikonal quasinormal modes} \\
\hline
Theory & Publication  \\[3pt]  \hline \\[-5pt]
Schwarzschild & \cite{MashhoonBlome} \\[5pt]
Schwarzschild-de Sitter  & \cite{Zhidenko:2003wq, Cardoso:2003sw}  \\[5pt]
Reissner-Nordstrom  & \cite{Andersson:1996xw}  \\[5pt]
Dilatonic black hole  & \cite{Konoplya:2001ji}  \\[5pt]
Kerr black hole & \cite{Mashhoon, Konoplya:2016pmh, Konoplya:2017tvu} \\[5pt]
Extremal Kerr & \cite{Hod:2012bw} \\[5pt]
Higher dimensional Schwarzschild & \cite{Konoplya:2003ii} \\[5pt]
Higher dimensional, extremal  Schwarzschild-de Sitter & \cite{Molina:2003ff} \\[5pt]
Higher dimensional charged Einstein-Gauss-Bonnet & \cite{Konoplya:2004xx} \\[5pt]
Einstein-dilaton-Gauss-Bonnet & \cite{Konoplya:2019hml} \\[5pt]
Einstein-Weyl & \cite{Kokkotas:2017zwt} \\[5pt]
Reissner-Nordstrom with derivative coupling to a scalar field & \cite{Konoplya:2018qov} \\[5pt]
Brane-world black holes & \cite{Kanti:2005xa, Zhidenko:2008fp} \\[5pt]
\hline
\end{tabular}
\end{table*}
\end{center}

\section{Spherically symmetric black holes spacetimes}

A static, spherically symmetric metric in the spacetimes under consideration has the form:
\begin{equation}\label{metric}
d s^2 = -f(r) d t^2 + \frac{1}{f(r)} d r^2 +r^2\left(\sin^2\theta d\phi^2+d\theta^2\right).
\end{equation}

The metric function is given by:
\begin{enumerate}
\item for the first kind Einstein aether BH
\begin{equation}\label{metricfunctionEA1}
f(r) = 1- \frac{2M}{r} - I\left(\frac{2M}{r}\right)^4, \; I=\frac{27c}{256(1-c)},
\end{equation}
\item and for the second kind Einstein aether BH
\begin{equation}\label{metricfunctionEA2}
f(r) = 1- \frac{2M}{r} - J\left(\frac{M}{r}\right)^2, \; J=\frac{c-d/2}{1-c},
\end{equation}
where $c$, $d$ are the combinations of the coupling constants of the theory, $0\leq c<1$, $0\leq d<2$, $c\geq d/2$ \cite{Ding:2017gfw}, \cite{ Ding:2019tvs}. Einstein aether theory is a Lorentz-violating theory endowing a spacetime with both a metric and a unit timelike vector field (aether) having a preferred time direction;

\item for the Horava-Lifshitz BH
\begin{equation}\label{metricfunctionHL}
f(r) = 1+w r^2 - \sqrt{r\left(w^2 r^3+4w M\right)},
\end{equation}
where $M$ is an integration constant, $w$ is a theory parameter, $w M^2\geq 1/2$ \cite{Kehagias:2009is}. Quasinormal modes were analyzed in \cite{Konoplya:2009ig, Chen:2009gsa};

\item for the Hayward BH  \cite{Hayward:2005gi}
\begin{equation}\label{metricfunctionH}
f(r) = 1- \frac{2Mr^2}{r^3+Q^3},
\end{equation}
where $Q$ is some real positive constant;

\item for the Bardeen BH \cite{Bardeen}
\begin{equation}\label{metricfunctionB}
f(r) = 1- \frac{2Mr^2}{\left(r^2+Q^2\right)^{3/2}},
\end{equation}
where $Q$ is a magnetic charge;

\item for the Ayon-Beato-Garcia (ABG) BH \cite{ABG}
\begin{equation}\label{metricfunctionABG}
f(r) = 1- \frac{2Mr^2}{\left(r^2+Q^2\right)^{3/2}}+\frac{Q^2r^2}{\left(r^2+Q^2\right)^{2}},
\end{equation}
where $Q$ is an electric (or magnetic) charge.
\end{enumerate}

\section{Eikonal expansion}

\begin{center}
\begin{table*}
\begin{tabular}{p{4cm}p{2cm}cc}
\multicolumn{4}{c}{{\bf Table 2.} Corrections for eikonal quasinormal modes in terms of the small parameters of the metrics} \\
\hline
Metric & $\alpha$ & $\Delta\omega_R$ & $\Delta\omega_I$  \\[3pt]  \hline \\[-5pt]
Einstein-aether I & $I$ & $-\frac{8}{27}\alpha+\frac{352}{729}\alpha^2+{\cal O}(\alpha^3)$ & $-\frac{16}{27}\alpha+\frac{1664}{729}\alpha^2+{\cal O}(\alpha^3)$ \\[5pt]
Einstein-aether II & $J$ &  $-\frac{1}{6}\alpha+\frac{13}{216}\alpha^2+{\cal O}(\alpha^3)$ & $\frac{1}{18}\alpha+\frac{1}{72}\alpha^2+{\cal O}(\alpha^3)$  \\[5pt]
Horava-Lifshitz & $\frac{1}{w M^2}$ & $\frac{1}{27}\alpha+{\cal O}\left(\alpha^2\right)$ & $\frac{2}{27}\alpha+{\cal O}\left(\alpha^2\right)$ \\[5pt]
Hayward & $\frac{Q}{M}$ & $\frac{1}{27}\alpha^3+{\cal O}\left(\alpha^4\right)$ & $\frac{2}{27}\alpha^3+{\cal O}\left(\alpha^4\right)$ \\[5pt]
Bardeen & $\frac{Q}{M}$ & $\frac{1}{6}\alpha^2+\frac{17}{216}\alpha^4+{\cal O}\left(\alpha^5\right)$ & $\frac{1}{9}\alpha^2+\frac{149}{648}\alpha^4+{\cal O}\left(\alpha^5\right)$  \\[5pt]
ABG & $\frac{Q}{M}$ & $\frac{1}{3}\alpha^2+{\cal O}\left(\alpha^3\right)$ & $\frac{1}{18}\alpha^2+{\cal O}\left(\alpha^3\right)$ \\[5pt]
\hline
\end{tabular}
\end{table*}
\end{center}

The perturbations of the black hole can be represented in the general form of the wave like equation
\begin{equation}\label{wave-equation}
\frac{d^2\Psi}{dr_*^2}+\left(\omega^2-V(r)\right)\Psi=0,
\end{equation}
where $r_*$ is the "tortoise coordinate", mapping the event horizon to $-\infty$,
\begin{equation}
dr_*=\frac{dr}{f(r)}.
\end{equation}
The boundary conditions for this equation are only incoming waves at the horizon ($r_*\rightarrow -\infty$) and only outgoing waves at the infinity($r_*\rightarrow +\infty$). Solving the wave equation we obtain a discrete set of complex values for the frequencies $\omega$ with the real part representing the oscillation frequency and imaginary part representing the damping rate of the oscillations in terms of the black hole parameters.

\begin{figure}[h!]
\includegraphics[width=1\linewidth]{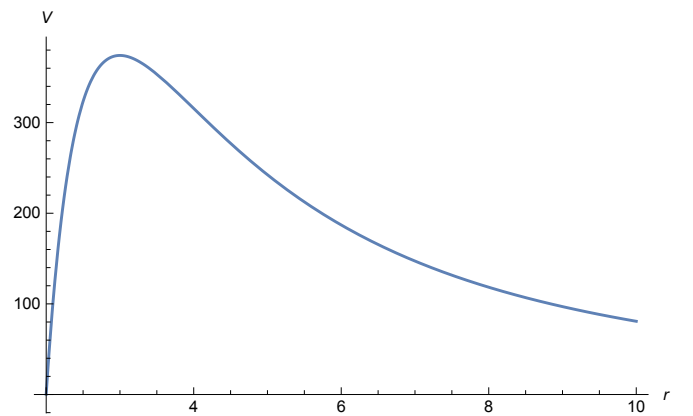}
\caption{An example of an effective potential: the Hayward BH in the eikonal regime ($\ell=100$, $\alpha=0.1$)}
\label{pot}
\end{figure}

To find a solution of Eq. (\ref{wave-equation}) we use the WKB approximation \cite{WillSchutz}:
\begin{equation}\label{WKB}
\omega^2=V_0+\sqrt{-2V''_0}\left(n+\frac{1}{2}\right)i,
\end{equation}
which is accurate in the eikonal limit and therefore produces analytical results.

The effective potential has the form of the potential barrier (see, for example Fig.\ref{pot})
and its maximum position can be expanded in terms of a small parameter $\alpha$:
\begin{equation}
r_{max}=r_0+r_1\alpha+r_2\alpha^2+r_3\alpha^3+{\cal O}(\alpha^4),
\end{equation}
where $r_0=3M$ corresponds to the Schwarzschild solution.

In the eikonal limit the effective potential usually does not depend on spin of the field, though there are a number of exceptions, connected either to existence of the eikonal instability or non-linear couplings. Here we will not consider such cases. Thus, the effective potential has the form:
\begin{equation}\label{potential}
V(r)=f(r)\left(\frac{\ell\left(\ell+1\right)}{r^2}+{\cal O}(1)\right),
\end{equation}
which implies applicability of the obtained analytical formulae for the quasinormal modes to all kinds of perturbations, such as gravitational, scalar, Dirac, etc., but except perturbations of non-linear electromagnetic fields in the background of charged black holes \cite{Toshmatov:2018ell, Toshmatov:2018tyo, Toshmatov:2019gxg}.

For each metric under consideration there is a number of reasons to consider additional parameters as small: these are either requirements of good post-Newtonian behavior or other astrophysical constraints. Therefore we find the eikonal quasinormal modes in the form
$$
\omega=\frac{1}{3\sqrt{3}M}\left(\ell+\frac{1}{2}\right)\left(1+\Delta\omega_R\right)-
$$

\begin{equation}\label{QNM}
-i\frac{1}{3\sqrt{3}M}\left(n+\frac{1}{2}\right)\left(1+\Delta\omega_I\right)+{\cal O}\left(\frac{1}{\ell +\frac{1}{2}}\right),
\end{equation}
where $\Delta\omega_R$ and $\Delta\omega_I$ are the corrections for the real and imaginary part of the quasinormal modes with regard to the Schwarzschild case.
The parameters and corrections are presented in Table 2.

Unlike all the cases studied above the Horndeski theory  \cite{Fang:2018vog} is characterized by the metric
$$
d s^2 = -\left(1-\frac{2M}{r}+\frac{c}{r^{4+\frac{1}{\alpha^2}}}\right) d t^2 +
$$
$$
+
\left(1-\frac{2M}{r}+\frac{c}{r^{4+\frac{1}{\alpha^2}}}\right)^{-1} d r^2+
$$
\begin{equation}\label{metricHr}
 +\frac{r^2}{\left(1+4 \alpha^2
 \right)}\left(\sin^2\theta d\phi^2+d\theta^2\right),
\end{equation}
which is not asimptotically flat. The effective potential for electromagnetic perturbations has the form  \cite{Fang:2018vog}
\begin{equation}\label{potentialHr} 
V(r)=\left(1-\frac{2M}{r}+\frac{c}{r^{4+\frac{1}{\alpha^2}}}\right)
\frac{\ell\left(\ell+1\right)\left(1+4\alpha^2\right)}{r^2}
\end{equation}
and the formula for the eikonal quasinormal modes in this theory reads
$$
\omega=\frac{1}{3\sqrt{3}M}\left(\ell+\frac{1}{2}\right)\sqrt{1+\frac{3c}{\left(3M\right)^K}} \; -
$$
$$
- \; i\frac{1}{3\sqrt{3}M}\left(n+\frac{1}{2}\right)\sqrt{1-\frac{cK^2}{2\left(3M\right)^K}-\frac{3c^2K^2}{\left(3M\right)^{2K}}} \; +
$$
\begin{equation}\label{QNMHr}
+ \; {\cal O}\left(\frac{1}{\ell +\frac{1}{2}}\right),
\end{equation}
where
$$
K=4+\frac{1}{\alpha^2},
$$
$c$ is a theory constant and $\alpha$ is a small parameter, such that when it tends to zero the metric approaches that of the Schwarzschild spacetime. Then, $K$ becomes infinitely large, and $\left(3M\right)^K$ is large too, while
$
K^2/(3M)^K
$
is small again. Thus when the theory parameter $\alpha$ tends to zero, the corrections in Eq. (\ref{QNMHr}) vanish and we come to the same Schwarzschild case.

\section{General approach for asymptotically flat metrics}

The idea of expanding the position of the peak of the effective potential in terms of a small parameter in order to present it in the form of the Schwarzschild peak with added relatively small corrections can be extended onto the metric function itself.
Having  an asymptotically flat metric we can expand it in terms of the negative powers of $r$:
\begin{equation}\label{expMetric}
f(r)=1-\frac{2M}{r}+\frac{\alpha_2}{r^2}+\frac{\alpha_3}{r^3}+\frac{\alpha_4}{r^4}+\frac{\alpha_5}{r^5}+\; {\cal O}\left(\frac{1}{r^6}\right),
\end{equation}
where parameters $\alpha_i$ describe deviations of a given metric from the Schwarzschild one. The maximum of the effective potential can be expanded as follows:
\begin{equation}
r_{max} = r_0+r_1\alpha_2 +r_2\alpha_3+r_3\alpha_4+r_4\alpha_5 +{\cal O}(\alpha_i \alpha_k).
\end{equation}
Using the first order WKB formula and expanding the results for $\omega$ into powers of $\alpha_i$, we find that 
$$
\omega=\frac{\left(\ell+\frac{1}{2}\right)}{3\sqrt{3}M}\left(1+\frac{\alpha_2}{6 M^2}+\frac{\alpha_3}{18 M^3}+\frac{\alpha_4}{54 M^4}+\frac{\alpha_5}{162 M^5}\right)-
$$

$$
-i\frac{\left(n+\frac{1}{2}\right)}{3\sqrt{3}M}\left(1+\frac{\alpha_2}{18 M^2}-\frac{\alpha_3}{27 M^3}-\frac{\alpha_4}{27 M^4}-\frac{11 \alpha_5}{486 M^5}\right)+
$$

\begin{equation}\label{QNMGen}
+ \; {\cal O}\left(\frac{1}{\ell +\frac{1}{2}}\right).
\end{equation}

The eikonal formulas, presented in Table 2, can then be immediately found from the above general formula by using the coefficients $\alpha_i$  obtained via expansion of the metric in terms of the corresponding small parameters of the system.

\section{Conclusions}

We filled the gap in the current literature devoted to analytical calculations of quasinormal modes in the high frequency regime. For the set of the spherically symmetric metrics we obtained corrections for the real and imaginary part of the eikonal quasinormal modes in terms of the small parameters of the theories. We also deduced a general formula for eikonal quasinormal modes for the class of the asymptotically flat spacetimes in terms of small deviations from Schwarzschild geometry. In addition we reviewed publications where analytical eikonal formula was derived for various black hole spacetimes and fields. We believe that this compendium of analytical formulae for high frequency regime of black hole perturbations might be of further usage  in analytic and semi-analytic treatments of black hole perturbations \cite{Hod:2012bw, Hod:2018ifz,  Zinhailo:2018ska}.

\acknowledgments{The author acknowledges hospitality and support of Silesian University in Opava, Roman Konoplya for suggesting this problem and useful discussions and Alexander Zhidenko and Bobir Toshmatov for valuable advice.}

\end{document}